\documentstyle[epsf,aps,prl,12pt,preprint]{revtex}
\def\baselinestretch{1.5}
\title{\bf {\em Minimum} Entropy Approach to Word Segmentation Problems}
\author{Bin Wang \\   
 {\small Institute of Theoretical Physics, Chinese Academy of 
Sciences},\\
  {\small P.O. Box 2735, Beijing 100080, P. R. China.}\\
 {\small State Key Laboratory of Scientific and Engineering Computing},\\
{\small Institute of Computational Mathematics and Scientific/Engineering Computing}, \\
  {\small P.O. Box 2719, Beijing 100080, P. R. China.}\\
  }
\begin{document}
\maketitle
%\widetext
\vspace {1cm}
\begin{abstract}
Given a sequence composed of a limit number of characters, 
 we try to ``read" 
 it as a ``text". This involves to segment the sequence into ``words". The 
difficulty is to distinguish  good segmentation from enormous number of 
random ones.
Aiming at revealing the nonrandomness of the sequence as strongly as 
possible, 
by applying maximum likelihood method, we find a 
quantity called {\bf segmentation entropy} that can be used to 
fulfill the duty. Contrary to commonplace 
where maximum entropy principle was applied to obtain good solution, we 
choose to {\em minimize} the segmentation entropy to obtain good 
segmentation. The concept developed in this letter can be used to 
study the noncoding DNA sequences, e.g., for regulatory elements 
prediction, in eukaryote genomes.

\vspace {0.4cm}
\noindent
%PACS number: 87.10.+e, 89.70.+c. 
\end{abstract}
%\widetext

\clearpage
\newpage
%\begin {multicols}{2}

\section{Introduction.}
The problem addressed in this paper is rather 
elementary in statistics. It is best described as the following: 
suppose one who knows nothing about English language 
was given a sequence of English letters, which was actually 
obtained by taking off all the interwords delimiters 
among a sample of English text,
how could he recover the words of the text by choosing to 
insert spaces between 
adjacent letters? Note that the only thing he can
consult is the statistical properties of the sequence?

Any two adjacent letters can be chosen to belong to the same word (keep 
adjacent) as 
well as belong to separate words (be separated by space). 
Suppose the sequence length is $N$. 
Any choice on the connectivity between $N-1$ pairs of 
adjacent letters is called a segmentation. 
There are a total of $2^{N-1}$ possible
segmentations. The word segmentation problem is to find ways to 
distinguish the 
correct segmentation -- in the sense that adjacent letters in the 
original text keep 
adjacent while letters separated by spaces and/or punctuation 
marks in the original 
text are separated by spaces in the segmentation -- from others. 

Although the problem seems toy-like, its fundamental 
importance for statistical linguistics is evident. We study on it, 
however, also for practical purposes. Noncoding sequences in the genomes 
of species play 
essential rule on the regulation of gene expression and function~\cite{liw}. 
However the development of computational methods for extracting regulatory elements is far behand DNA sequencing and gene finding~\cite{regulate}.
One reason is the lack of efficient way to discriminate large 
amount of sequence signals in noncoding DNA sequences. 
Through linguistic study it has been shown that noncoding sequences in 
eukaryotic genomes are structurally much similar to natural and 
artificial language~\cite{stanley}. Thus many may expect to ``read" the 
noncoding sequences as a ``text". Actually, efforts have been given to 
build a dictionary for genomes~\cite{trifonov,li}. Li et al.~\cite{li} 
showed the connection between regulatory elements prediction 
and word segmentation in noncoding DNA sequences of eukaryote genomes. 
We expect that 
progress on word segmentation problem may help to deepen our knowledge on 
noncoding 
regions of eukaryote genomes. Besides, word segmentation is an important 
issue for Asian languages (e.g., Chinese and Japanese) processing~\cite{ponte}, 
because they lack interword delimiters.

\section{Segmentation entropy and its connection to word segmentation problem.}
To tackle word segmentation problem,
we first consider a problem under constraints, so that one important 
concept --
 segmentation entropy -- can be introduced. The 
constraints will be released at the end of this paper. 
Suppose we have known that there are $n_l$ words of length $l$ 
$(l=1,2,\cdots)$ in the original text. Obviously,
\begin {equation}
\sum_l{n_ll}=N. 
\end {equation}
Under these constraints -- Words Length Constraints WLC -- there are 
totally 
\begin {equation}
\frac{(\sum_l{n_l})!}{\prod_l{n_l}!}
\end {equation}
segmentations. For example, for the following story, there are totally 
$3.12e144$ segmentations, while the number under WLC is about $1.33e97$. 

\begin {quote}
\begin {center}
{\sl
                       The Fox and the Grapes
}
\end {center}
{\small
{\sl
   \ \ \ \ Once upon a time there was a fox strolling through the woods. 
 He came upon a grape orchard.  There he found a bunch of beautiful 
grapes hanging from a high branch.	

  \ \ \ \ ``Boy those sure would be tasty," he thought to himself.  He 
backed up and took a running start, and jumped.  He did not get high 
enough.

  \ \ \ \ He went back to his starting spot and tried again.  He almost 
got high enough this time, but not quite.	

  \ \ \ \ He tried and tried, again and again, but just couldn't get 
high enough to grab the grapes.	

  \ \ \ \ Finally, he gave up.	

  \ \ \ \ As he walked away, he put his nose in the air and said: ``I am 
sure those grapes are sour."
}
}
\end {quote}

Following least effort principle~\cite{zipf}, it is appreciable in 
natural languages to 
combine existing words to express different meaning. 
Shannon~\cite{shannon} pointed out the 
importance of redundancy in natural languages long ago: generally 
speaking, nearly half of the letters 
in a sample of English text can be deleted while someone else can still 
restore them. These 
properties of natural language ensure the sequence obtained by 
taking off interword delimiters from a certain text being highly nonrandom 
and showing determinant and regular characteristics. 
It is expected that the correct segmentation reveals these 
characteristics as 
strongly as possible. From information point of view, this means 
that, if 
a form of information entropy can be properly defined on each segmentation, 
the entropy of the correct 
segmentation will be the smallest.

Interestingly, a maximum likelihood approach leads to the same 
proposal and automatically gives the definition of the entropy.
Given one sequence of length $N$, we expect to find a likelihood function which 
reaches its maximum on the correct 
segmentation. For a concrete segmentation, we assign a probability to 
each word in it 
\begin {equation}
w_i\to{p_i}, \qquad  i=1...M
\end {equation}
with
\begin {equation}
\sum_{i=1}^M{p_i}=1.
\end {equation}
The likelihood function is written as
\begin {equation}
Z_s=\prod_{i=1}^M{{p_i}^{m_il_i}}
\end {equation}
where $m_i$ is the number of word $w_i$ in the segmentation, and $l_i$ 
is the length of the word.

By maximizing the likelihood function subjected to 
eq.(4) we obtain
\begin {equation}
p_i=\frac{m_il_i}{N}.
\end {equation}
Thus the maximum likelihood for the segmentation is
\begin {equation}
Z_s=\prod_{i=1}^M{(\frac{m_il_i}{N})^{m_il_i}}.
\end {equation}
The segmentation with maximum likelihood is just the one minimizing 
\begin {equation}
S=-\frac{lnZ_s}{N}=-\sum_{i=1}^M{\frac{m_il_i}{N}ln(\frac{m_il_i}{N})}.
\end {equation}
This function has the form of entropy~\cite{shannon} and will be called Segmentation Entropy (SE).

Starting from a maximum likelihood approach, we now come to the 
suggestion 
to minimize the segmentation entropy. 
This is in contrast to commonplace. 
Maximizing likelihood leads to maximizing certain entropy in some 
cases~\cite{frieden,jaynes}. 
As a general principle for investigating statistical problems, maximum 
entropy method has been successfully applied in a 
variety of fields~\cite{frieden,jaynes}. We propose that, instead of 
applying maximum entropy principle, one may choose to minimize certain entropy 
(minimum entropy principle) in some problems. This seems attractive 
especially in the era of bioinformatics when most of the 
problems are to reveal regularity in large amount of seemingly 
random sequences.

Because the present is a statistical method, the 
text under study needs to be not too short. For example, when we 
tried to 
find the segmentation with the smallest segmentation entropy for the 
saying 
\begin {quote}
{\sl God is nowhere as much as he is in the soul... and the soul 
means the world} 
\end {quote}
(By Meister Eckhart, 14-century Dominican 
priest, Preacher, and Theologian), it was found that, among a total of 
$343062720$ segmentations under WLC, 
there are 15 segmentations 
whose SE is $2.3684$, smaller than 2.3802 of the correct one. One example is 
\begin {quote}
{\sl god isnow {\bf he} rea smuchas {\bf he} is int {\bf he} {\bf soul} andt 
{\bf he} {\bf soul} meanst {\bf he} world}, 
\end {quote}
in which the five {\em ``he"} and two {\em ``soul"} are 
revealed.

Unfortunately, present computational power does not permit to 
exhaustively 
study even a text as short as {\sl ``the Fox and the Grapes"}, the 
number of 
permitted segmentations for which is $1.33E+97$ under WLC. 
We choose to see the relevance of the concept of segmentation entropy 
in some special ways. The study focuses on ``The Fox and the Grapes".

To change a segmentation slightly, one way is to choose two adjacent words 
along the sequences randomly and then exchange their length. This way the 
original two words may change to different words. This procedure 
can be repeated on the resulting segmentations. The change does not violate the WLC. Because of the large number of possible choices in each step, the segmentation is expected to become increasingly dissimilar to the original one. Starting from the correct segmentation of ``The Fox and the Grapes", we expect to see the evolution of SE by changing the segmentation this way.
Figure 1 shows that SE increase drastically in the first 500 steps, 
and then reaches and fluctuates around certain equilibrium value. 
Compared with the gap between the equilibrium value and the original  
SE, the fluctuation is minor.
This shows that, at least locally, the correct segmentation is at the 
minimum of 
SE. Actually, we have traced a trajectory of evolution up to $10^{10}$ 
steps. No 
segmentation with SE smaller than the correct one was observed. This 
implies that SE of the correct segmentation is also globally minimal.

The distribution of segmentation entropy may give 
further insight to the atypicality of 
the correct SE. 
We randomly sampled $10^{10}$ segmentations in the following way: while 
keeping the WLC, the length of each words in the segmentation is assigned 
randomly. The distribution of SE is shown in Fig. 2. 
The minimal SE we sampled is 4.5298, still much higher 
than 4.097 of the correct segmentation (see Fig. 1). 
It is interesting to observe that the 
distribution shows fractal characteristics. The fractal-like distribution 
presents also for 
other text, even for random sequence (Fig. 3). The fractal-like feature 
is determined by the WLC and the statistical structure of the sequence 
under study.
In Fig. 3 we compared the distribution of SE of two 
sequences (under the same WLC), the original sequence of {\sl ``The Fox and the Grapes"} and 
a random sequence obtained by randomizing the order of letters in the text.
The result is in accordance with the fact that the original 
sequence is in a much more ordered state, manifesting that 
segmentation entropy captures the statistical structure of the 
sequences successfully.

There is one way to estimate the number of segmentations the SE of which is 
4.097, the value for the correct segmentation.
See Fig. 4 in which the distribution of SE in Fig. 2 are shown 
in logrithmic scale here. The left edge of the distribution fall on a line. 
The edge can be fitted by $e^{(165x-750.42)}.$ 
The number of segmentations with SE x among the totally $1.33e97$ possible 
segmentations under WLC is:
\begin {equation}
c(x)=\frac{1.33e^{97}}{9\times10^9}e^{(165x-750.42)}.
\end {equation}
We obtained $c(4.097)=0.96$. 
From the distribution of SE shown in Fig. 3(a) we obtained the same value of $c(4.097)$. The estimation support the idea that segmentation entropy of 
correct segmentation is unique.

We now consider how to release the WLC. 
Unfortunately, searching the 
segmentation with the smallest SE among all the possible 
is sure to fail to find the correct one. For example, SE of the 
segmentation in which the whole sequence 
is considered as one word (single-word segmentation) 
is 0, the smallest possible SE. 
Also, the 
segmentation in which each letter is viewed as a separate word 
($N$-word segmentation) has a considerably small 
SE (2.8655 for {\sl ``The fox and the grapes"}). 
These are called side attraction effects. These examples show that smaller 
SE does not necessarily means better segmentation 
when we compare the SEs of segmentations under 
different WLC (here WLC refers to any partition of numbers of 
words of various length 
satisfying eq.(1), not necessarily the same as the original text.)
The bias induced by different WLC must be taken off.
In order to do so, we suggest to use 
\begin {equation}
R_S=\frac{S}{S_0}
\end {equation}
instead of $S$.
Here $S_0$ is the average SE under the same WLC of a sequence obtained 
by randomizing the order of letters in the original text.
$S_0$ plays the role of chemical potential for a thermodynamic system~\cite{chempot}.
$R_S$ for the single word and $N$-word segmentations are 1, the largest 
possible value.
By searching segmentation with the smallest $R_S$, it is expected to 
find meaningful segmentation. For examples, for the segmentation 
\begin {quote}
{\sl god isnow {\em he} rea smuchas {\em he} is int {\em he} {\em soul} andt {\em he} {\em soul} meanst {\em he} 
world,}
\end {quote}
which has already been shown above, $R_S$ 
is 0.8601; while 
\begin {quote}
{\sl god {\em is} now {\em he} re {\em as} much {\em as} {\em he} {\em is} int {\em he} {\em soul} {\em an} dt {\em he} {\em soul} me {\em an} 
st {\em he} world} 
\end {quote}
is a better -- actually one of the best -- segmentation according to 
$R_S$ ($R_S=0.8259$). Intuitively this is reasonable, because in this  
second segmentation, more repeated ``words" -- two copies of 
{\em ``is"}, {\em ``as"} and {\em ``an"} -- are revealed. 
Another segmentation 
\begin {quote}
{\sl god {\em is} now {\em he} re {\em as} much {\em as} {\em he} {\em is} in {\em thesoul} {\em an} d {\em thesoul} me {\em an} st 
{\em he} world}, 
\end {quote}
which differs from the second segmentation by revealing the two 
{\em ``thesoul"}, has a moderately small $R_S$: 0.8481. 
Comparison shows that the five repeats of 
{\em ``he"} is the most preferred part in good segmentations.

\section {Concluding remarks.}
In statistical linguistics many efforts are given on 
signal extracting and statistical inference. 
Our method, however, is new on at least two points. First, there is neither 
assumption on distribution~\cite{peitra} nor demand for training 
sets, lexical or grammatical knowledge~\cite{ponte}. 
This feature is important for studying biological 
sequences, because present knowledge on the ``language" (DNA) 
of life is still lack.
Second, instead of extracting a limit number of signals, 
we try to ``read" the sequence exactly as a ``text". 
A text includes more than words: it also includes the organization of words.
The results of segmentation form a basis for many further elaborations.

Principally, the concept of segmentation entropy can be applied to study 
the noncoding DNA sequences of eukaryote genomes. It is expected that the 
study may gives more than some meaningful ``words" or regulatory 
elements. Possible applications are not 
 confined to studying noncoding DNA sequences of course. Segmentation 
entropy can be used to find patterns in any symbolic sequences. 
However,
the application of segmentation entropy is restricted by the difficulty to find 
the segmentation with the smallest $R_s$ from the vast amount possible 
ones. We are now developing algorithm that can be used for regulatory binding sites prediction. in the algorithm the principle of minimun entropy will be incorporated in. 

\section*{ACKNOWLEDGMENTS}
I thanks Professor Bai-lin Hao who helps to make 
the computing possible. I also thanks Professor Wei-mou Zheng and 
Professor Bai-lin Hao for stimulating discussions. Mr. Xiong Zhang carefully 
read the manuscript. The work was supported 
partly by National Science Fundation.

\clearpage
\newpage

\begin{figure}[p]
\vspace {2cm}
\centerline{\epsfxsize=10cm \epsfbox{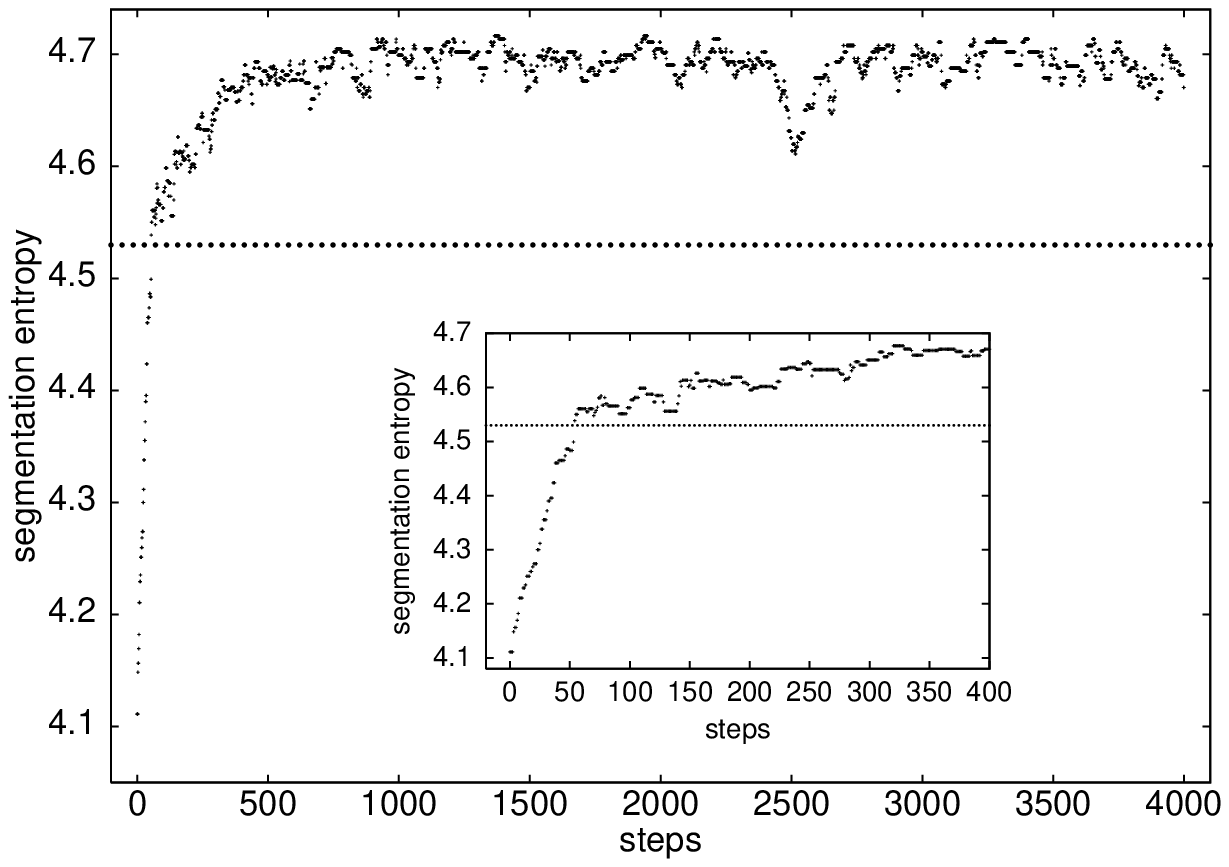}}
\label{evolve}
\vspace {2cm}
\caption{The evolution of segmentation entropy. Starting from the 
correct one, the segmentation was change stepwisely by 
exchanging the lengths of a pair of adjacent words randomly chosen along the 
sequence. 
The doted line corresponds to the smallest segmentation entropy 4.5298 
for the $10^{10}$ randomly sampled segmentations, see Fig. 2.}
\end{figure}

\clearpage
\newpage

\begin{figure}[p]
\vspace {2cm}
\centerline{\epsfxsize=10cm \epsfbox{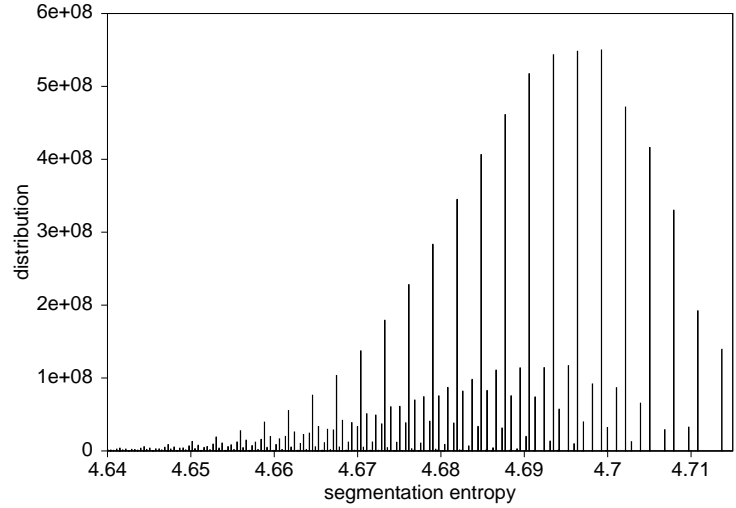}}
\label{stat}
\vspace {2cm}
\caption{The distribution of the segmentation entropy of 
$9\times10^{9}$ segmentations randomly chosen for the text ``The Fox 
and the Grapes". The numbers of words of various length in the original 
text were first counted. In the sampled segmentations these numbers were 
kept, but the length of each word along the sequence were randomly 
assigned.}
\end{figure}

\clearpage
\newpage

\begin{figure}[p]
\vspace {2cm}
\centerline{\epsfxsize=10cm \epsfbox{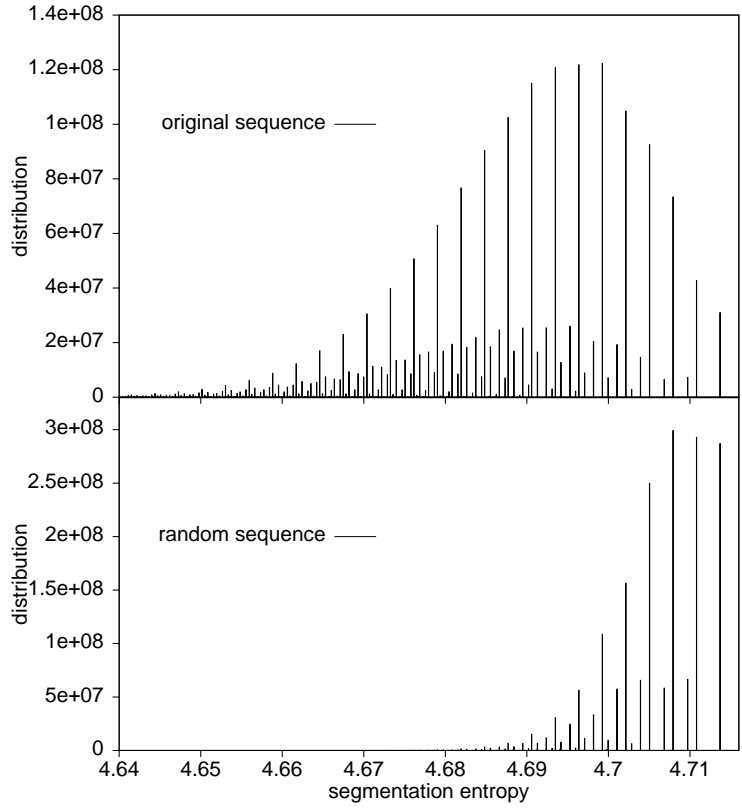}}
\label{compare}
\vspace {2cm}
\caption{Comparison of the distribution of segmentation entropy for two 
sequences: the original sequence of {\sl ``The Fox and the Grapes"}, and 
a random sequence obtained by randomizing the order of letters in the 
original text. For each sequence, $10^{9}$ segmentations are randomly 
sampled in the way described in the caption of Fig. 2.}
\end{figure}

\clearpage
\newpage

\begin{figure}[p]
\vspace {2cm}
\centerline{\epsfxsize=10cm \epsfbox{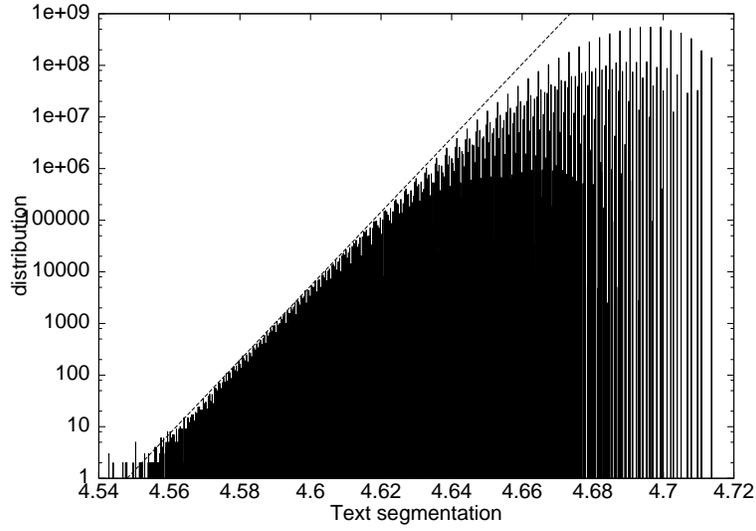}}
\label{fit}
\vspace {2cm}
\caption{The distribution of segmentation shown in Fig. 2 is shown in log 
scale here. The line along the left edge of the distribution is 
$e^{(165x-750.42)}$.}
\end{figure}
%\end{multicols}

\end{document}